\documentclass[aps,prb,showpacs,twocolumn,superscriptaddress]{revtex4}
\usepackage{amsmath,graphics}
\usepackage[next]{inputenc}
\usepackage[dvips]{epsfig}
\usepackage{hyperref}
\def\be{\begin{equation}}
\def\ee{\end{equation}}

\def\bi{\begin{itemize}}
\def\ei{\end{itemize}}
\def\bn{\begin{enumerate}}
\def\en{\end{enumerate}}
\def\bea{\begin{eqnarray}}
\def\eea{\end{eqnarray}}

\def\ba{\begin{array}}
\def\ea{\end{array}}
\def\bd{\begin{displaymath}}
\def\ed{\end{displaymath}}

\begin{document}
\title{Universality and Scaling Properties of Correlation Functions Near a Quantum
Phase Transition}

\author{R. Jafari}
\affiliation{Research Department, Nanosolar System Company (NSS), Zanjan 45158-65911, Iran}
\email[]{jafari@iasbs.ac.ir, jafari@nss.co.ir}
\affiliation{Department of Physics, Institute for Advanced
Studies in Basic Sciences (IASBS), Zanjan 45137-66731, Iran}

\begin{abstract}
In this paper we investigate the universality and scaling properties of the well-known quantities in
classical statistical mechanics near the quantum phase transition point.
We show that transverse susceptibility and derivatives of correlation functions with respect to the parameter that drives the quantum phase transitions, exhibit logarithmic divergence and finite size scaling the same as entanglement.
In other words the non-analytic and finite size scaling behaviors of entanglement
is not its intrinsic properties and inherit from the non-analytic and scaling behaviors of
correlation functions and surveying at least the nearest neighbor correlation functions could specify the scaling
and divergence properties of entanglement. However we show that the correlation functions could capture the quantum critical point without pre-assumed order parameters even for the cases where the two-body entanglement is absent.

\end{abstract}
\date{\today}


\maketitle
\section{Introduction \label{introduction}}
Quantum Phase Transition (QPT) is a continuous phase transition which occurs at zero
temperature where, quantum fluctuations play the dominant role.
Suppression of the thermal fluctuations at zero temperature introduces
the ground state as the representative of the system.
Often the abrupt changes in the ground state manifest itself in the non-analytic behavior of
macroscopic properties of the system. But in a number of important cases this criterion fails to predict a QPT, such as the three-dimensional metalinsulator transition of non-interacting electrons in a random potential \cite{Edwards}.
The QPT are typically accompanied by certain exponents pertaining to the class of quantum phase transitions \cite{Sachdev}. The concept of correlation, i.e. information of one part of a system
about another part, is a key element in (QPT) physics. Correlations can emerge from classical and quantum sources.

The existence of genuinely quantum correlations can be usually inferred by the presence of entanglement
among parts of a system. Indeed, entanglement displays a rather interesting behavior at QPT, being able to indicate a Quantum Critical Point (QCP) through nonanalyticities \cite{Osterloh,Shi,Amico1,Kargarian,Jafari1,Jafari}.
Moreover, for one-dimensional critical systems, ground state entanglement entropy exhibits a universal logarithmic
scaling \cite{Korepin}. In several years an idea has arranged that a complete classification of the critical
many-body state requires the introduction of concepts from quantum information theory \cite{Osterloh,Shi}.

Although it is known that the ground state and correlation functions could capture the QCP, but to the best of our
knowledge, universality and scaling properties of well-known correlation functions have not been investigated. In this work, we will study the universality and scaling properties of correlation functions and the Transverse
Susceptibility (TS) to compare with universality and scaling properties of entanglement.
We show that derivative of correlation functions with respect to the transverse field,
exhibit a logarithmic singularity at the QCP and show the scaling behavior
close to the QCP the same as TS. The scaling behavior and constants which are obtained in this work are the same as
that obtained in refs. [\onlinecite{Osterloh}] and [\onlinecite{Shi}] for the entanglement.
In other words, divergence of entanglement close the QCP is not surprising since the reduced density matrix and consequently the two-body entanglement are smooth functions of correlation functions and
its non-analytic behavior and finite size scaling rise out of divergence
and scaling behavior of correlation functions' derivative.
However, we show that the correlation functions between far neighbors are able to
characterize a QPT, even for distances where pairwise entanglement is absent.
This is a consequence of the longer range of correlation functions at the critical point in
comparison with the short-range behavior of pairwise entanglement.

\section{XY Model in Transverse Field and Nearest-Neighbor correlation functions}

The system under consideration is the antiferromagnetic spin-1/2 $XY$ chain
in a transverse magnetic field, a model central both to condensed-matter
and information theory and subject to intense study \cite{Brooke}.

The hamiltonian is
\be
\label{eq1}
H=\sum_{i=1}^{N}[J_{x}\sigma_{i}^{x}\sigma_{i+1}^{x}+
J_{y}\sigma_{i}^{y}\sigma_{i+1}^{y}+h\sigma_{i}^{z}],
\ee

where $J_{x}>0$ and $J_{y}=>0$ are the exchange couplings,
$h$ is the transverse field and $N$ is the number of sites.
This model is exactly solvable \cite{Pfeuty} with
the Jordan-Wigner transformation which in the momentum space leads to
$H=\sum_{k}2\epsilon_{k}(\gamma^{\dag}_{k}\gamma_{k}-\frac{1}{2})$,
where $\gamma^{\dag}_{k} (\gamma_{k})$ is the fermion creation (annihilation)
operator. For states of even fermions,
\be
\label{eq2}
\epsilon_{k}=\sqrt{[h+(J_{x}+J_{y})\cos(k)]^{2}+(J_{x}-J_{y})^{2}\sin^{2}(k)},
\ee
and the ground state energy is given by $E_{G}=-\sum_{k}\epsilon_{k}$,

with $k=\frac{2n\pi}{N}, n=-\frac{N}{2},\cdots,\frac{N}{2}-1$.
For $J_{y}=0$ equation (\ref{eq1}) reduces to the Ising model in transverse field (ITF), whereas for
$h=0$ it is the XY model. For all the interval $0<J_{x}+J_{y}\leq 1$ the models belong
to the Ising universality class and for $N=\infty$ they undergo a quantum phase
transition at the critical value $\lambda_{c}=(J_{x}+J_{y})/h=1$. The magnetization $<\sigma^{x}>$
is different from zero for $\lambda>1$ and it vanishes at the transition point.
On the contrary the transverse magnetization $<\sigma^{z}>$ is different from
zero for any value of $\lambda$. At the phase transition the correlation length
$\xi$ diverges as $\xi\sim|\lambda_{c}-\lambda|^{-\nu}$, with $\nu=1$ \cite{Pfeuty}.

The two point nearest neighbor (NN) correlation functions and transverse
magnetization are coincidently the expectation value of the coupling term
in the Hamiltonian \cite{Barouch},

\be
\label{eq3}
G^{xx}=\frac{dE_{g}}{dJ_{x}},~G^{yy}=\frac{dE_{g}}{dJ_{y}},~<\sigma^{z}>=\frac{dE_{g}}{dh}.
\ee
The non-analytic behavior and finite-size scaling of two-body NN entanglement
of ITF model have been studied intensively \cite{Osterloh,Shi}. Then first, we look at the ITF case
($J_{x}=1,J_{y}=0$). In order to quantify the connection between the non-analytic behavior of Entanglement
when the system crosses the critical point and statistical mechanics,
we look at the derivative of the NN correlation functions and TS as a function of $h$
(for simplicity we have consider the absolute value of the ground state energy).

\begin{figure}
\begin{center}
\includegraphics[width=9cm]{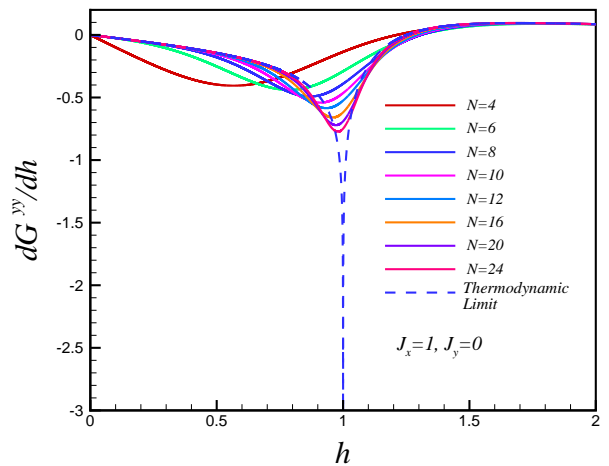}
\caption{(Color online) Evolution of the $\frac{dG^{yy}}{dh}$ versus $h$ for
different system sizes for $J_{x}=1, J_{y}=0$.} \label{fig1}
\end{center}
\end{figure}

In Fig.{\ref{fig1}} we have shown the derivative of $G^{yy}$ with respect to the magnetic field
versus $h$.

\begin{figure}
\begin{center}
\includegraphics[width=9cm]{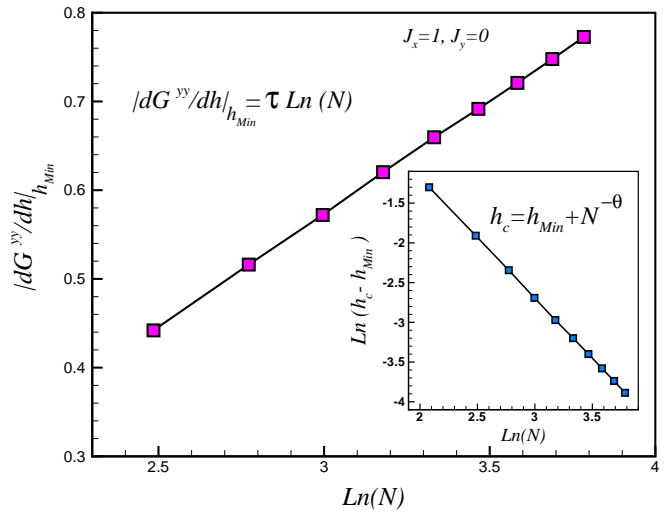}
\caption{(Color online) Scaling of the minimum of $\frac{dG^{yy}}{dh}$ for
systems of various sizes. Inset: Scaling of the position ($h_{Min}$) of $\frac{dG^{yy}}{dh}$
for different-length chains.} \label{fig2}
\end{center}
\end{figure}

As expected, there is no divergence for finite lattice sizes, but
for the infinite chain $dG^{yy}/dh$ diverges on approaching the critical point.
A more detailed analysis shows that the position of the minimum ($h_{Min}$) of $dG^{yy}/dh$
tends towards the critical point as $h_{Min}=h_{c}-N^{-\theta}$	with $\theta=1.50\pm0.03$
(Fig.(\ref{fig2}), inset). Moreover, we have derived the scaling behavior of $|dG^{yy}/dh|_{h_{Min}}$
versus N. This is plotted in Fig. (\ref{fig2}), which shows the linear behavior of $|dG^{yy}/dh|_{h_{Min}}$
versus ln(N). The scaling behavior is $|dG^{yy}/dh|_{h_{Min}}=\tau\ln(N)$ with
$\tau=0.25\pm0.03$. It is interesting that it is very close to the ITF
correlation function exponent $\eta=\frac{1}{4}$.

\begin{figure}[b]
\begin{center}
\includegraphics[width=9cm]{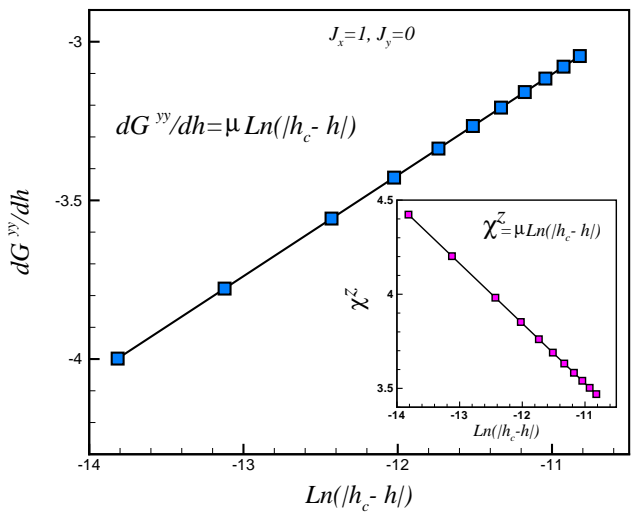}
\caption{(Color online) Logarithmic divergence of $|\frac{dG^{yy}}{dh}|$
close the critical point for the infinite system. Inset: Logarithmic divergence of
transverse susceptibility ($\chi^{z}$) near the critical point for the infinite
system.}\label{fig3}
\end{center}
\end{figure}

For an infinite system, $dG^{yy}/dh$  diverges on approaching the critical point as
$\frac{dG^{yy}}{dh}=\mu\ln(|h_{c}-h|)$, where shown in Fig.(\ref{fig3}).
Here $\mu$ is $0.27\pm0.03$ but the analytical calculation
shows that it is not completely independent of $h$.

According to the scaling ansatz \cite{Barber}, the two-site correlation functions,
considered as a function of the system size and the coupling, is a function of $N^{1/\nu}(h-h_{Min})$.
In the case of logarithmic divergence which obtained in Fig.\ref{fig3}, it behaves as
$(dG^{yy}/dh-dG^{yy}/dh|_{h_{Min}})\sim F(N^{1/\nu}(h-h_{Min}))$.

\begin{figure}[t]
\begin{center}
\includegraphics[width=9cm]{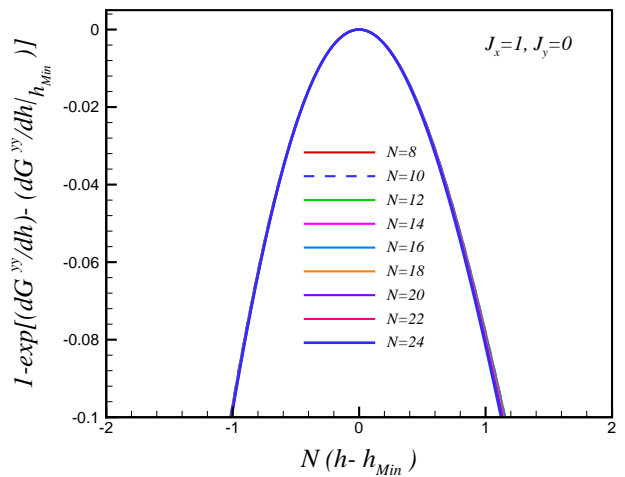}
\caption{(Color online) Finite-size scaling of $dG^{yy}/dh$
for different lattice sizes. The curves which correspond to
different system sizes clearly collapse on a single curve.}
\label{fig4}
\end{center}
\end{figure}

In this way we have plotted $1-exp(dG^{yy}/dh-dG^{yy}/dh|_{h_{Max}})$
versus $N^{1/\nu}(h-h_{Min})$ in Fig.(\ref{fig4}). The curves which correspond
to different system sizes clearly collapse on a single universal curve.
Our result shows that $\nu=1\pm0.001$ is exactly correspond to the correlation length
exponent of ITF ($\nu=1$).

A similar analysis can be carried on $G^{xx}$, $G^{zz}$ \cite{Barouch} and TS ($\chi^{z}$).
Our calculations show that the non-analytic and scaling behavior of
NN correlation functions and TS are the same as each other. The logarithmic divergence of
TS near the critical point has been shown in inset of Fig.(\ref{fig3}) in which
$\mu=0.27\pm0.03$ equals the scaling coefficient of $dG^{yy}/dh$.
In order to confirm our findings, we investigate analytically the derivative of correlation
functions and TS using the exact ground state expression of ITF model which is obtained using the
Eq.(\ref{eq2}) for $J_{x}=1, J_{y}=0$.
It is not hard to see that the ground state energy per site can be written as
a complete elliptic integral of the second kind \cite{Pfeuty}.
\be
\label{eq5}
E_{G}=-\frac{2J_{x}(1+\gamma)}{\pi}E(\frac{\pi}{2},a),
\ee
in which $a=\frac{2\sqrt{\gamma}}{1+\gamma}$ and
$\gamma=\frac{1}{\lambda}=\frac{h}{J_{x}}$.
In the vicinity of $a=1$, $E(\frac{\pi}{2},a)\sim 1+\frac{1-a^{2}}{4}
\ln\frac{16}{1-a^{2}}$ \cite{Abramowitz}, and therefore in the vicinity of critical
point ($\lambda_{c}=h_{c}=\gamma=1$) the ground state energy contain the nonanalytic
contribution $E_{G}=(h-h_{c})^{2}\ln|h-h_{c}|$.

\begin{figure}[t]
\begin{center}
\includegraphics[width=9cm]{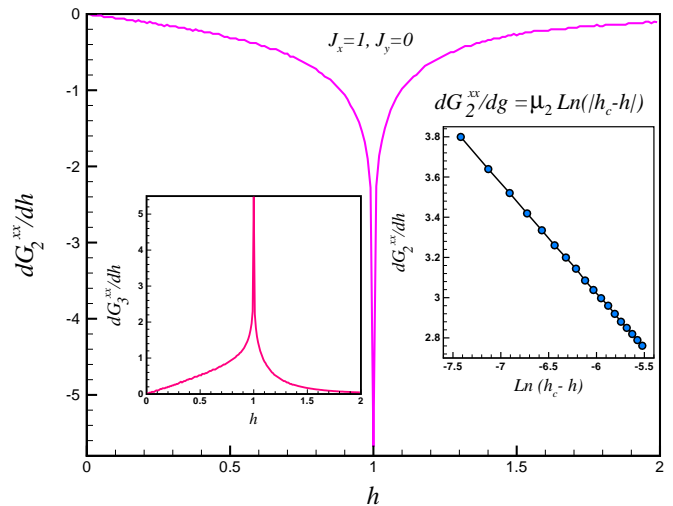}
\caption{(Color online) Derivative of next nearest neighbor
correlation function ($G^{xx}_{2}$) versus h for infinite size system.
Right inset: Logarithmic divergence of next nearest neighbor correlation
function near the critical point for the infinite size system. Left inset:
Derivative of third nearest neighbor correlation function ($G^{xx}_{3}$)
versus h for the infinite size system.}
\label{fig5}
\end{center}
\end{figure}

Similarly the $G^{xx}$ correlation function and transverse
magnetization ($<\sigma^{z}>$), which are obtained using the
derivative of ground state with respect to the $J_{x}$ and the magnetic field, could be written
as a functions of complete elliptic integrals of first and second kinds \cite{Abramowitz},
\be
\label{eq6}
G^{xx}=-\frac{1}{2\pi}\Big[\frac{2(4+\lambda)}{1+\lambda}K(\frac{\pi}{2},a)-
\frac{8}{\sqrt{\lambda}}\frac{1}{da}E(\frac{\pi}{2},a)\Big]
\ee
where $K(\frac{\pi}{2},a)$ is complete elliptic integrals of first kind.
Then after some complicated calculations, we could obtain $\frac{G^{xx}}{dh}\sim\frac{G^{yy}}{dh}\sim\chi^{z}\sim\mu\ln|h-h_{c}|$ in which the coefficient of $\mu$ is function of $h$.

\section{Second and Third Nearest Neighbor correlation functions and their Scalings}

The numerical results shows that the first derivative of next nearest neighbor
(NNN) correlation functions with respect to the magnetic field, diverge at the
critical point (Fig.(\ref{fig5})) while the second derivative of NNN concurrence presents
such non-analyticity \cite{Osterloh}. This difference could originate
from the properties of the entanglement. The entanglement is linear function of the correlation
functions and shows the collective behavior of them in which the details of
the subsystem's behavior could be hidden.
the logarithmic divergence of NNN correlation functions has been shown
in right inset of Fig.(\ref{fig5}) with $\mu_{2}=0.51\pm0.03$.
Despite the discrepancy, both of NNN concurrence and NNN correlation functions
manifest the logarithmic divergence and same finite size scaling ($\nu=1$).
Moreover the third NN correlation functions's derivative diverge at the critical point
(Fig(\ref{fig5}), left inset) and their finite size scaling behaviors are the same as
NN and NNN correlation functions. It is important to mention that even when entanglement
is not present (for ITF case entanglement is indeed completely absent for sites farther
than second nearest-neighbors \cite{Osterloh}), the QPT can be clearly revealed through
the singular behavior of correlation functions.

\begin{figure}[t]
\begin{center}
\includegraphics[width=9cm]{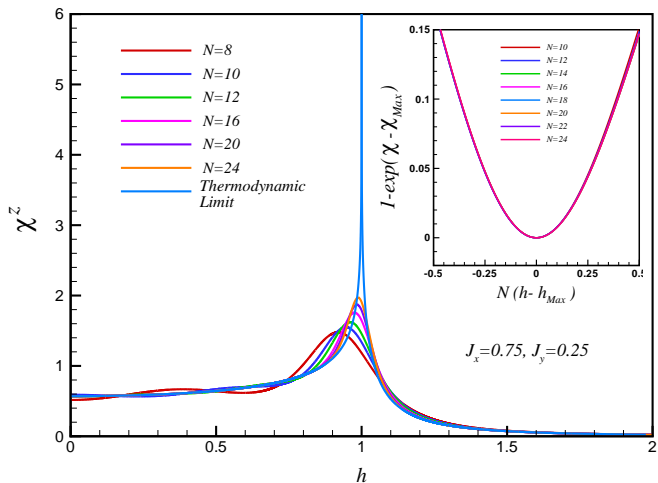}
\caption{(Color online) Evolution of the transverse susceptibility versus $h$ for
different system sizes for $J_{x}=0.75, J_{y}=0.25$. Inset: Finite-size scaling of $\chi^{z}$
for different lattice sizes. The curves corresponding to different lattice
sizes collapse to a single graph.} \label{fig6}
\end{center}
\end{figure}

\begin{figure}[t]
\begin{center}
\includegraphics[width=9cm]{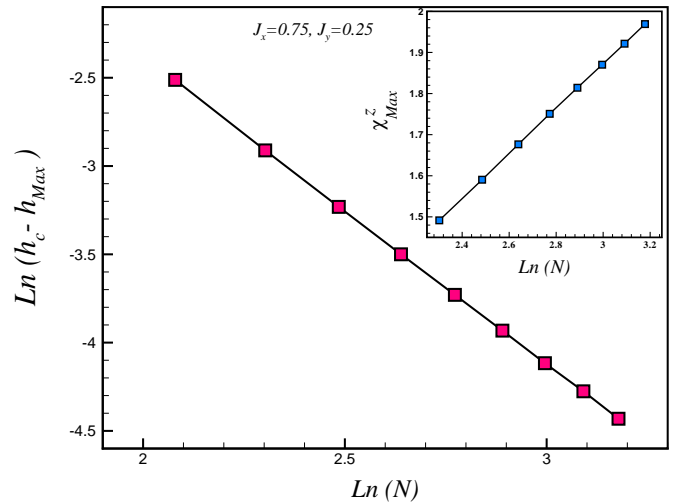}
\caption{(Color online)  Scaling of the position ($h_{Max}$) of $\chi^{z}$
for different-length chains. Inset:Scaling of the maximum of $\chi^{z}$ for
different lattice size.} \label{fig7}
\end{center}
\end{figure}

Second, we consider the case for $J_{y}\neq0$. Fig.(\ref{fig6}) specifies the divergence and
finite size scaling  of TS (Fig(\ref{fig6}), inset) for $J_{x}=0.75, J_{y}=0.25$  (correspond to the case $\gamma=0.5$ in ref. \cite{Osterloh}). In this case scaling is fulfilled with the critical exponent $\nu=1$ in agreement with the previous results and universality hypothesis. However the scaling of the maximum's position ($h_{Max}$) of TS has been shown in Fig.(\ref{fig7}) which manifest the linear behavior of $\ln(h_{c}-h_{Max})$ versus $\ln(N)$. Inset of Fig.(\ref{fig7}) shows the scaling of the maximum of $\chi^{z}$ versus the system size.

We have also studied the correlation functions and TS of quantum compass model \cite{Eriksson} in the presence of
homogenous and inhomogeneous magnetic field. This model shows a rich phase diagram which includes several critical point depending on the exchange coupling \cite{Jafari2}. In this model the correlation functions of the various
component of the two spins on odd bond are different from those one on even bond. Therefor this model is reasonable model to examine the above results. We have been able to obtain the divergence and scaling behaviors of the
NN correlation functions and TS of this model \cite{Jafari2}. This results justify the above finding.

\section{Summary and Discussion}

To summarize, we have investigate the properties of the correlation functions and we have shown that because of their nice scaling properties, phase transition point can be determined from small systems with considerable accuracy without pre-assumed order parameters even for the cases where the pairwise entanglement is absent. Then any smooth function of correlation functions (concurrence, quantum discord, reduced density matrix) could bear the same dominating long range physics about the phase transition as correlation functions do.
However, we show that the peak anomaly of NN concurrence is exactly dominated by short range interactions of the Hamiltonian. So studying the divergence and scaling properties of correlation functions enable us to study the none-analytic and scaling properties of entanglement without direct calculation of entanglement (concurrence) which is elaborate calculation. However, for some cases in the one dimension, the correlation functions' predicting power of phase transition owe to their long-range behaviors could be more reliable than same predicting by the entanglement where shows short-range behavior and divergence in its second order derivative.
Moreover, our results suggest that because of nice scaling properties of correlation functions, they are proper quantities to obtained the QCP from the finite size data where definition of a suitable order parameter (susceptibility) is difficult.

Further investigations including blocks and multi-body correlations functions may be interesting to establish a precise comparison between universality and scaling behaviors of correlation functions and entanglement at critical points especially in higher dimension. Such topics are left for a future research.

\begin{acknowledgments}
The author would like to thank \textbf{V. Karimipour}, A. Langari,
M. R. .Kolahchi, R. Fazio, S. Mahdavifar, M. Abedi and M. Nasiri for reading the manuscript,
fruitful discussions and comments.
\end{acknowledgments}


\section*{References}
\begin{thebibliography}{99}

\bibitem{Edwards}
J. T. Edwards, and D. J. Thouless, Regularity of density of states in Andersons localized electron
model. J. Phys. C \textbf{4}, 453 (1971).

\bibitem{Sachdev}
S. Sachdev, \emph{Quantum Phase Transitions} (Cambridge University
Press, Cambridge, 2000).

\bibitem{Osterloh}
A. Osterloh, Luigi Amico, G. Falci and Rosario Fazio, Nature
\textbf{416}, 608 (2002).

\bibitem{Shi}
Shi-Quan Su, Jun-Liang Song, and Shi-Jian Gu, Phys. Rev. A \textbf{74},
032308 (2006).

\bibitem{Amico1}
L. Amico, R. Fazio, A. Osterloh, and V. Vedral, Rev. Mod. Phys.
\textbf{80}, 517 (2008).

\bibitem{Kargarian}
M. Kargarian, R. Jafari, and A. Langari, Phys. Rev. A \textbf{76}, 060304(R) (2007).


\bibitem{Jafari1}
R. Jafari, M. Kargarian, A. Langari, and M. Siahatgar, Phys. Rev. B \textbf{78}, 214414 (2008).

\bibitem{Jafari}
R. Jafari, Phys. Rev. A \textbf{82}, 052317 (2010).

\bibitem{Korepin}
V. E. Korepin, Phys. Rev. Lett. \textbf{92}, 096402 (2004).

\bibitem{Brooke}
J. Brooke, , D. Bitko, T. F. Rosenbaum, and G. Aeppli, Quantum annealing
of a disordered spin system. Science \textbf{284}, 779-781 (1999).

\bibitem{Pfeuty}
P. Pfeuty, ANNALS of Physics, \textbf{57}, 79 (1970).

\bibitem{Barouch}
E. Barouch, and B. M. McCoy, Phys. Rev. A \textbf{3}, 786 (1971).

\bibitem{Barber}
M. N. Barber, in Phase Transitions and Critical Phenomena
(Academic, London, 1983), Vol. \textbf{8}, pp. 146-259.


\bibitem{Abramowitz}
M. Abramowitz and I. A. Stegun, Eds., Handbook of Mathematical Functions
with Formulas, Graphs and Mathematical Tables (National Bureau of Standards, 1964).


\bibitem{Eriksson}
E. Eriksson and H. Johannesson, Phys. Rev. B \textbf{79}, 224424 (2009).


\bibitem{Jafari2}
R. Jafari, e-print arXiv:1101.3673v1.

\end{thebibliography}
\end{document}